# A Multi-agent Simulation for the Mass School Shootings

Wei Dai*, Senior Member, IEEE, Yash Pratap Singh †, Rui Zhang ‡
*,‡ Department of Computer Science,  † Department of Electrical and Computer Engineering
*Purdue University Northwest*
Hammond, Indiana, USA
*{\*weidai,  † sing1625,  ‡ zhan5097}@pnw.edu*

*Abstract*—The increasing frequency of mass school shootings in the United States has been raised as a critical concern. Active shooters kill innocent students and educators in schools. These tragic events highlight the urgent need for effective strategies to minimize casualties. This study aims to address the challenge of simulating and assessing potential mitigation measures by developing a multi-agent simulation model. Our model is designed to estimate casualty rates and evacuation efficiency during active shooter scenarios within school buildings. The simulation evaluates the impact of a gun detection system on safety outcomes. By simulating school shooting incidents with and without this system, we observe a significant improvement in evacuation rates, which increased from 16.6% to 66.6%. Furthermore, the Gun Detection System reduced the average casualty rate from 24.0% to 12.2% within a period of six minutes, based on a simulated environment with 100 students. We conducted a total of 48 simulations across three different floor layouts, varying the number of students and time intervals to assess the system's adaptability. We anticipate that the research will provide a starting point for demonstrating that a gunshot detection system can significantly improve both evacuation rates and casualty reduction.

## I. INTRODUCTION

Mass school shootings have become a pressing global issue, raising concerns over the safety of educational institutions [1]. In [2], the authors analyzed mass school shooting data in the USA from 1999 to 2024. There are 43 mass school shootings with average 21 casualties per incident. The frequency of these tragic events has surged in recent years, necessitating a deeper understanding of their underlying causes and the development of effective strategies to minimize casualties. These incidents not only result in the loss of lives but also generate long-lasting psychological trauma for survivors and communities [3]. Thus, improving safety measures during active shooting situations is of paramount importance.

While various studies have sought to analyze the factors contributing to mass school shootings, many have highlighted the significant challenges associated with collecting reliable, comprehensive data on these highly sensitive events. The majority of available data, particularly on the dynamics of shootings within school environments, remain sparse and incomplete [4]. However, emerging technologies such as simulation modeling offer a promising avenue for addressing these limitations. By creating virtual environments that mimic real-world conditions, simulations allow for detailed analysis of critical factors such as evacuation routes, response times, and casualty rates during an active shooting. The main contributions of our work are:

*(a)* : We have unified the actions of three agents in the agent-based simulation by employing mathematical formulas that ensure proper movement dynamics, enhancing the realism and effectiveness of agent interactions within the environment.

*(b)* : We have conducted a comprehensive comparison between the gun detection system and the non-gun detection system, providing insights into their respective impacts on agent behavior and overall evacuation efficiency in critical scenarios.

The rest of this paper is structured as follows: Section II highlights the previous work in the domain, Section III outlines the methodology of the approach. Section IV discusses the results of the study in detail and provides an in-depth analysis of the findings. Finally, Section V is the conclusion and future work.

## II. RELATED WORK

Simulations are powerful research approaches with broad applications in aerospace engineering [5], nuclear research [6], quantum computing [7], fluid dynamics [8], network security [9], and time synchronization [10]. Using simulation to model complex systems is cost-effective, enabling the optimization of strategies, prediction of outcomes, risk reduction, and efficient resource allocation. Simulations enable dynamic evaluations, enhance efficiency, and support safer, data-driven solutions [11], making them invaluable for addressing challenges such as increasing school safety.

In [2], the authors introduced two mathematical models based on game theory. The research split a mass school shooting event into four parts: The killers identify victims (KIV) phase, the Victims Attacked (VA) phase, the Police Officers Moving (POM) phase, and the Shootout phase. A Miller curve describes the accumulated loss function, *Loss*(*t*), and the injury function, *I*(*t*) in four stages. The Miller curve

The Provost Office of Purdue University Northwest and the Indiana Space Grant Consortium funded the research project.

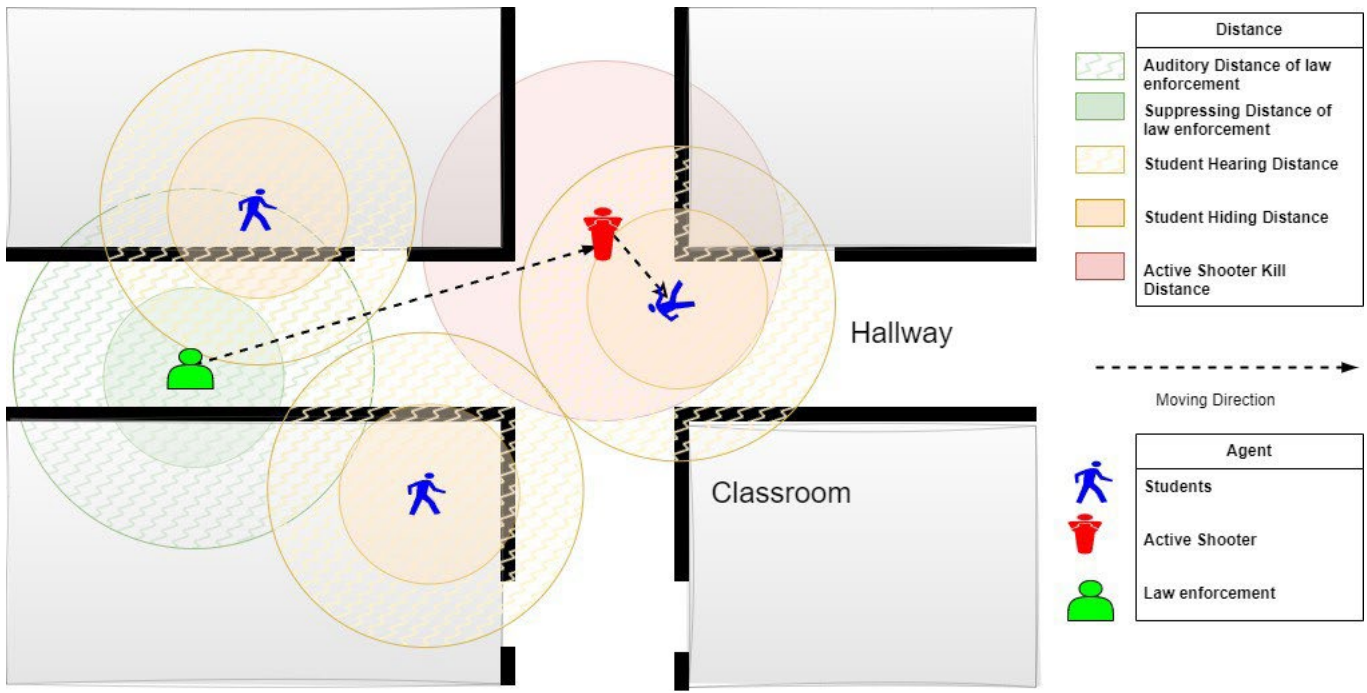

Fig. 1. Three agents in a building simulation, including law enforcement, students, and active shooter

provides theoretical and practical insight to improve school safety policies.

The work by Arteaga and Park [12] demonstrated a systematic evaluation of three key building design parameters - building exit width, door width, and hall width - to assess their potential effects on reducing casualties and improving evacuation efficiency during indoor mass school shootings. Using agent-based modeling and the social forces model of pedestrian dynamics, they simulated 150 scenarios in a school-like building setting at various occupancy levels. Their results indicated that hall and door widths had the most significant observed benefits for improving victim safety. Their findings provide a foundation of analytical evidence to initiate discussions about the importance of considering active shooter scenarios when designing facilities. To make the results more realistic, Cotfas et al. [13] used agent-based modeling (ABM) to mimic human behavior during emergencies and studied evacuation procedures in major event halls. Their methodology improved the realism of evacuation simulations by accounting for varied agent characteristics, including age, speed, and impairments. The study's "adapted cone exit" strategy beat conventional techniques by directing agents to the nearest exit. Six scenarios were simulated, and the results showed that ineffective departure options and crowding were the leading causes of evacuation delays. Event coordinators and safety engineers can learn much from this effort, especially regarding large-scale evacuations. The potential of machine learning (ML) to improve ABM in complex systems was thoroughly reviewed by Zhang et al. [14]. Their research revealed four key areas in which ML might support agent-based behavior modeling: enhancing agents' context awareness, modifying agent behaviors, simulating macro-level ABM outcomes, and streamlining decision-making processes. Specifically, the incorporation of reinforcement learning demonstrated promise for enhancing the flexibility and judgment of agents. With the help of this multidisciplinary framework, ABM has more applications in a wider range of disciplines, including environmental management and sociology.

While the studies mentioned above have significantly contributed to understanding evacuation dynamics through Agent-Based Simulations, they fall short of proposing a practical, implementable solution for improving evacuation efficiency in real-world scenarios. Arteaga and Park [12], as well as Cotfas et al. [13], focus on how structural parameters like hall width, exit width, and door size influence casualty rates and evacuation times. However, retrofitting existing school buildings to meet these ideal structural conditions is often impractical and cost-prohibitive. Additionally, these studies do not account for the immediate threat detection and response mechanisms that can critically alter the course of an active shooter situation. To mitigate these issues, our work addresses this gap by evaluating the integration of a gun detection system with evacuation protocols, offering a solution that can be deployed in existing facilities without major structural alterations. By detecting threats in real-time and dynamically adjusting evacuation strategies based on the shooter's location and movement, we can significantly improve evacuation efficiency and reduce casualties. This technology-driven approach provides a more feasible, scalable alternative to relying solely on modifying building layouts, making it a practical solution for enhancing safety in schools and other public spaces.

## III. METHODOLOGY

Due to the impracticality of conducting real-life drills or simulations to study victim behavior during a mass shooting, ABM [15] emerges as a highly viable alternative for addressing this issue through computational simulation. This section further details the *(i)* Algorithm for students, shooter, and law-enforcement movement, *(ii)* Structures of different floors, and *(iii)* Simulation setup.

TABLE I
EXPERIMENTAL SETUP FOR DIFFERENT FLOOR STRUCTURES

|  | No. of Experiments Performed | No. of Rooms | Total Exits |
|---|---|---|---|
| Structure 1 | 16 | 23 | 4 |
| Gyte (Floor 1) | 16 | 30 | 5 |
| SULB (Floor 3) | 16 | 27 | 3 |

### A. Agents

In the simulation, agents are the core components that drive the model's dynamics, each playing a distinct role based on predefined behaviors and objectives. The primary agents in this scenario are the students, the shooter, and the law enforcement officer, all of which are designed with unique behavioral patterns to mimic real-world actions in a school shooter scenario. All agent's behavior and movement under different conditions are illustrated in Fig. 1.

*1) Student's Behavior:* The simulated behavior of the students follows 3 types of movements: run, hide, or no effect. The behavior and movement patterns of students vary significantly depending on the presence or absence of a Gun Detection System on the same floor. When a Gun Detection System is in place, all students are alerted after the first shot, making them aware of the shooter's location. As a result, the majority of students evacuate the building safely, while those in close proximity to the shooter choose to hide. Conversely, in the absence of such a system, students do not receive a notification after the first shot. So, those situated further from the incident will not hear the shot and thus remain unaware of the threat, while students within audible range will start to run despite not knowing the shooter's exact location. Additionally, students in close proximity choose to hide. This variation in behavior under the two conditions significantly impacts evacuation efficiency and casualty rates. The given equations can represent all the student's actions on a particular map structure.

For Non-Gun Detection system floor :

$$\text{Actions S} = (f_{non}(dist(i, cfk, cs)))_{i=1}^{N} \quad (1)$$

For Gun Detection system floor :

$$\text{Actions S} = (f_{gun}(dist(i, cfk)))_{i=1}^{N} \quad (2)$$

where cfk = coordinates of first kill, cs = coordinates of shooter, and N = total number of students

The function $f_{non}$ and $f_{run}$ are calculated using the below equations, in this $\alpha$ represents hiding range and $\beta$ represents hearing range:

$$f_{non} = H \cdot \theta(\alpha - dist_{fk}) + R \cdot \theta(dist_{sh} - \alpha) \cdot \theta(\beta - dist_{sh}) \quad (3)$$

where H: Hide, and R: Run. The function $\theta(x)$ represents the Heaviside step function, which is 1 if $x \geq 0$ and 0 otherwise.

$$\theta(\alpha - dist_{fk})$$

ensures that the distance for hiding is within range.

$$\theta(dist_{sh} - \alpha) \cdot \theta(\beta - dist_{sh})$$

ensures that the distance is between the hiding and hearing ranges.

$$f_{gun} = H \cdot \Theta(\alpha - dist) + R \cdot \Theta(dist - \alpha) \quad (4)$$

$$\theta(\alpha - dist)$$

ensures that the distance for hiding is within range.

$$\theta(dist - \alpha)$$

ensures the condition when dist is out of hiding range. The dist is calculated using the Manhattan distance formula: For Non-Gun Detection system floor :

$$\text{dist} = \begin{cases} |X_i - X_{fk}| + |Y_i - Y_{fk}| & (dist_{fk}), \\ |X_i - X_{sh}| + |Y_i - Y_{sh}| & (dist_{sh}) \end{cases} \quad (5)$$

For Gun Detection system floor :

$$dist = (|X_i - X_{fk}| + |Y_i - Y_{fk}|) \quad (6)$$

*2) Shooter's Behavior:* The shooter navigates through the hallways in search of students, with a specified range within which they can target individuals. However, the probability of a successful kill diminishes as the distance between the shooter and the target increases. The shooter employs a ray-casting approach to detect targets within a 360-degree field of view. Upon identifying a student within range, the shooter attempts to eliminate them using a probability density function that reduces the likelihood of success as the distance grows. This process is repeated consistently at each step along the shooter's path until the simulation concludes at predetermined intervals *(6 mins, 7 mins, 8 mins, and 9 mins)*. The below equation represents the shooter's action; in the below equation, $\gamma$ represents the visible range of the shooter, K represents Kill, M represents Move, and $\hat{D}_{shooter-student}$ represents the unit vector pointing from the shooter to the nearest student.

$$\text{Actions K} = K \cdot \theta(\gamma - (student)_{i=1}^{N}) + M \cdot min(\hat{D}_{shooter-student}) \quad (7)$$

TABLE II
COMPARISON OF RESULTS WITH AND WITHOUT GUN-DETECTION SYSTEM FOR STRUCTURE 1

|  | Result 1 (Without Gun-Detection System) | | Result 2 (With Gun-Detection System) | | Comparison | |
|---|---|---|---|---|---|---|
|  | Casualties (%) | Evacuation (%) | Casualties (%) | Evacuation (%) | Casualty Change | Evacuation Efficiency Change |
| 50 Students | 19.60 | 13.20 | 16.80 | 60.00 | -2.80 | 46.80 |
| 100 Students | 24.00 | 16.60 | 12.20 | 66.60 | -11.80 | 50.00 |
| 150 Students | 13.06 | 22.26 | 10.66 | 72.00 | -2.4 | 49.74 |
| 200 Students | 25.50 | 9.50 | 15.80 | 63.80 | -9.70 | 54.30 |

*Simulation runtime = 6 minutes*

TABLE III
COMPARISON OF RESULTS WITH AND WITHOUT GUN-DETECTION SYSTEM FOR STRUCTURE 1

|  | Result 1 (Without Gun-Detection System) | | Result 2 (With Gun-Detection System) | | Comparison | |
|---|---|---|---|---|---|---|
|  | Casualties (%) | Evacuation (%) | Casualties (%) | Evacuation (%) | Casualty Change | Evacuation Efficiency Change |
| 50 Students | 20.80 | 15.60 | 16.80 | 68.40 | -4.00 | 52.80 |
| 100 Students | 26.60 | 18.60 | 12.80 | 70.00 | -13.80 | 51.40 |
| 150 Students | 15.33 | 24.93 | 10.66 | 74.40 | -4.67 | 49.47 |
| 200 Students | 29.00 | 12.40 | 15.80 | 67.50 | -13.20 | 55.10 |

*Simulation runtime = 7 minutes*

TABLE IV
COMPARISON OF RESULTS WITH AND WITHOUT GUN-DETECTION SYSTEM FOR STRUCTURE 1

|  | Result 1 (Without Gun-Detection System) | | Result 2 (With Gun-Detection System) | | Comparison | |
|---|---|---|---|---|---|---|
|  | Casualties (%) | Evacuation (%) | Casualties (%) | Evacuation (%) | Casualty Change | Evacuation Efficiency Change |
| 50 Students | 24.4 | 18.00 | 16.8 | 72.00 | -7.60 | 54.00 |
| 100 Students | 28.6 | 21.20 | 13 | 74.80 | -15.60 | 53.60 |
| 150 Students | 18.00 | 26.53 | 10.80 | 76.13 | -7.20 | 49.60 |
| 200 Students | 30.00 | 17.00 | 17.40 | 70.2 | -12.60 | 53.20 |

*Simulation runtime = 8 minutes*

TABLE V
COMPARISON OF RESULTS WITH AND WITHOUT GUN-DETECTION SYSTEM FOR STRUCTURE 1

|  | Result 1 (Without Gun-Detection System) | | Result 2 (With Gun-Detection System) | | Comparison | |
|---|---|---|---|---|---|---|
|  | Casualties (%) | Evacuation (%) | Casualties (%) | Evacuation (%) | Casualty Change | Evacuation Efficiency Change |
| 50 Students | 29.60 | 21.20 | 16.80 | 74.40 | -12.80 | 53.20 |
| 100 Students | 30.00 | 26.00 | 13 | 77.80 | -17.00 | 51.80 |
| 150 Students | 19.20 | 27.60 | 10.80 | 77.90 | -8.40 | 50.30 |
| 200 Students | 35.40 | 23.00 | 17.40 | 73.70 | -18 | 50.70 |

*Simulation runtime = 9 minutes*

ensures that the shooter kills if the student comes within the visible range of the shooter. The shooter wants to maximize casualties while also minimizing the time and distance. The optimized equation of the shooter is represented as:

$$K(k_i, t, d) = \arg\max_k \sum_{i=1}^{N} k_i \cdot \max\left(0, 1 - \frac{d}{\gamma}\right) \cdot \theta(\gamma - (student)_{i=1}^N)) - \arg\min(t + d) \quad (8)$$

where:
- $\sum_{i=1}^{N} k_i \cdot P(d_i) \cdot \theta(\gamma - (student)_{i=1}^N)$:
  Maximize the total number of kills.
- $\theta(\gamma - (student)_{i=1}^N)$ ensures that only students within the visible range $\gamma$ are targeted.

Constraint:
- $\arg\min(t + d)$:
  The shooter seeks to minimize the total time (t) and distance traveled (d).

B. Law-Enforcement's behavior:

The law officer enters the building 5 minutes after the simulation begins, facilitated by the Gun-Detection System's ability to quickly provide critical information. Upon entry, the officer starts by moving through the hallways, following a trajectory that avoids revisiting previously patrolled areas whenever possible. As he patrols, he actively searches for the shooter within his range. If the shooter is detected, the officer suppresses or neutralizes the threat. In addition to shooter suppression, the officer checks for students in distress. If any students are found within his range, he evacuates them immediately, ensuring their safety. The below equation represents the officer's action; in the below equation, $\Gamma$ represents the visible range of the officer, $\sigma$ represents the hearing range of the officer, S represents suppress, E represents evacuate, M represents Move, and $\hat{D}_{officer-shooter}$ represents the unit vector in the direction from the officer to the shooter.

$$\text{Actions } L = M \cdot \min(\hat{D}_{officer-shooter}) + S \cdot \theta(\Gamma - shooter) + E \cdot \theta(\sigma - (student)_{i=1}^N) \quad (9)$$

TABLE VI
COMPARISON OF RESULTS WITH AND WITHOUT GUN-DETECTION SYSTEM FOR GYTE (FLOOR 1)

|  | Result 1 (Without Gun-Detection System) | | Result 2 (With Gun-Detection System) | | Comparison | |
| --- | --- | --- | --- | --- | --- | --- |
|  | Casualties (%) | Evacuation (%) | Casualties (%) | Evacuation (%) | Casualty Change | Evacuation Efficiency Change |
| 50 Students | 19.60 | 20.40 | 16.80 | 71.20 | -2.80 | 50.80 |
| 100 Students | 32.20 | 24.80 | 26.60 | 64.00 | -5.60 | 39.20 |
| 150 Students | 23.59 | 19.86 | 22.13 | 70.40 | -1.46 | 50.54 |
| 200 Students | 29.40 | 25.20 | 27.00 | 63.00 | -2.40 | 37.80 |

\* Simulation runtime = 6 minutes

TABLE VII
COMPARISON OF RESULTS WITH AND WITHOUT GUN-DETECTION SYSTEM FOR GYTE (FLOOR 1)

|  | Result 1 (Without Gun-Detection System) | | Result 2 (With Gun-Detection System) | | Comparison | |
| --- | --- | --- | --- | --- | --- | --- |
|  | Casualties (%) | Evacuation (%) | Casualties (%) | Evacuation (%) | Casualty Change | Evacuation Efficiency Change |
| 50 Students | 25.60 | 22.00 | 16.80 | 74.80 | -8.80 | 52.80 |
| 100 Students | 35.40 | 28.80 | 26.80 | 65.20 | -8.60 | 36.40 |
| 150 Students | 26.40 | 22.79 | 22.13 | 73.73 | -4.27 | 50.94 |
| 200 Students | 31.50 | 28.70 | 27.10 | 66.10 | -4.40 | 37.40 |

\* Simulation runtime = 7 minutes

TABLE VIII
COMPARISON OF RESULTS WITH AND WITHOUT GUN-DETECTION SYSTEM FOR GYTE (FLOOR 1)

|  | Result 1 (Without Gun-Detection System) | | Result 2 (With Gun-Detection System) | | Comparison | |
| --- | --- | --- | --- | --- | --- | --- |
|  | Casualties (%) | Evacuation (%) | Casualties (%) | Evacuation (%) | Casualty Change | Evacuation Efficiency Change |
| 50 Students | 26.40 | 28.00 | 16.80 | 76.80 | -9.60 | 48.80 |
| 100 Students | 37.00 | 31.00 | 26.80 | 66.60 | -10.20 | 35.60 |
| 150 Students | 27.46 | 23.99 | 22.13 | 74.26 | -5.33 | 50.27 |
| 200 Students | 35.20 | 30.20 | 27.10 | 68.90 | -8.10 | 38.70 |

\* Simulation runtime = 8 minutes

TABLE IX
COMPARISON OF RESULTS WITH AND WITHOUT GUN-DETECTION SYSTEM FOR GYTE (FLOOR 1)

|  | Result 1 (Without Gun-Detection System) | | Result 2 (With Gun-Detection System) | | Comparison | |
| --- | --- | --- | --- | --- | --- | --- |
|  | Casualties (%) | Evacuation (%) | Casualties (%) | Evacuation (%) | Casualty Change | Evacuation Efficiency Change |
| 50 Students | 28.00 | 30.00 | 16.80 | 77.60 | -11.20 | 47.60 |
| 100 Students | 37.60 | 32.00 | 26.80 | 68.00 | -10.80 | 36.00 |
| 150 Students | 29.46 | 24.53 | 22.13 | 75.06 | -7.33 | 50.53 |
| 200 Students | 38.20 | 31.20 | 27.10 | 70.40 | -11.10 | 39.20 |

\* Simulation runtime = 9 minutes

ensures that the officer suppresses the shooter if he is within visible range of the officer and evacuates the students if the student comes within the hearing range of the officer. The officer's behavior focuses on maximizing safety and student evacuations while minimizing exposure to risk:

$$L(S_{self}, E, R) = \arg\max_{S_{self}, E} \left( S_{self} \cdot \theta(\Gamma - \text{shooter}) + \sum_{i=1}^{N} E_i \cdot \theta(\sigma - \langle \text{student} \rangle_{i=1}^{N}) - \arg\min \frac{t_{exposure}}{d_{\text{officer-shooter}} + \epsilon} \right)$$ (10)

where:

- $S_{self} \cdot \theta(\Gamma - \text{shooter})$:
  Maximize personal safety by ensuring the shooter stays out of the officer's range $\Gamma$.

- $\sum_{i=1}^{N} E_i \cdot \theta(\sigma - \langle \text{student} \rangle_{i=1}^{N})$:
  Maximize the number of students evacuated, considering only those within the hearing range $\sigma$.

Constraint:

- $\arg\min \frac{t_{exposure}}{d_{\text{officer-shooter}} + \epsilon}$
  The officer aims to minimize risk, where $d_{officer-shooter}$ is the distance between the officer and the shooter, $t_{exposure}$ is the time spent within the dangerous range, and $\epsilon$ is a small positive constant to prevent division by zero.

C. Floor Layout

To check the adaptability of our Gun-Detection System, we simulated 3 different building designs resembling a school facility, as illustrated in figures 5, 6, and 7 respectively. All the structures have varying numbers of exits and rooms of varying sizes. A descriptive view of the properties of different floor structures is given in Table I. At the start of each simulation, agents were randomly distributed across the available rooms.

TABLE X
COMPARISON OF RESULTS WITH AND WITHOUT GUN-DETECTION SYSTEM FOR SULB (FLOOR 3)

|  | Result 1 (Without Gun-Detection System) | | Result 2 (With Gun-Detection System) | | Comparison | |
| --- | --- | --- | --- | --- | --- | --- |
|  | Casualties (%) | Evacuation (%) | Casualties (%) | Evacuation (%) | Casualty Change | Evacuation Efficiency Change |
| **50 Students** | 15.60 | 0.40 | 10.40 | 50.00 | -5.20 | 49.60 |
| **100 Students** | 16.60 | 2.40 | 11.40 | 45.00 | -5.20 | 42.60 |
| **150 Students** | 13.20 | 3.73 | 9.73 | 42.53 | -3.47 | 38.80 |
| **200 Students** | 18.50 | 6.10 | 11.00 | 39.00 | -7.5 | 32.90 |

*\* Simulation runtime = 6 minutes*

TABLE XI
COMPARISON OF RESULTS WITH AND WITHOUT GUN-DETECTION SYSTEM FOR SULB (FLOOR 3)

|  | Result 1 (Without Gun-Detection System) | | Result 2 (With Gun-Detection System) | | Comparison | |
| --- | --- | --- | --- | --- | --- | --- |
|  | Casualties (%) | Evacuation (%) | Casualties (%) | Evacuation (%) | Casualty Change | Evacuation Efficiency Change |
| **50 Students** | 16.40 | 0.80 | 10.80 | 58.40 | -5.60 | 57.60 |
| **100 Students** | 18.60 | 2.80 | 11.40 | 53.00 | -7.20 | 50.20 |
| **150 Students** | 16.66 | 4.80 | 10.13 | 52.66 | -6.53 | 47.86 |
| **200 Students** | 21.60 | 8.30 | 11.30 | 45.30 | -10.3 | 37.0 |

*\* Simulation runtime = 7 minutes*

TABLE XII
COMPARISON OF RESULTS WITH AND WITHOUT GUN-DETECTION SYSTEM FOR SULB (FLOOR 3)

|  | Result 1 (Without Gun-Detection System) | | Result 2 (With Gun-Detection System) | | Comparison | |
| --- | --- | --- | --- | --- | --- | --- |
|  | Casualties (%) | Evacuation (%) | Casualties (%) | Evacuation (%) | Casualty Change | Evacuation Efficiency Change |
| **50 Students** | 20.00 | 3.20 | 11.20 | 65.20 | -8.80 | 62.00 |
| **100 Students** | 21.80 | 4.00 | 11.40 | 60.20 | -10.40 | 56.20 |
| **150 Students** | 18.40 | 7.19 | 10.13 | 57.86 | -8.27 | 50.67 |
| **200 Students** | 24.50 | 9.00 | 11.30 | 53.80 | -13.20 | 44.80 |

*\* Simulation runtime = 8 minutes*

TABLE XIII
COMPARISON OF RESULTS WITH AND WITHOUT GUN-DETECTION SYSTEM FOR SULB (FLOOR 3)

|  | Result 1 (Without Gun-Detection System) | | Result 2 (With Gun-Detection System) | | Comparison | |
| --- | --- | --- | --- | --- | --- | --- |
|  | Casualties (%) | Evacuation (%) | Casualties (%) | Evacuation (%) | Casualty Change | Evacuation Efficiency Change |
| **50 Students** | 21.60 | 4.00 | 12.40 | 67.60 | -9.20 | 63.60 |
| **100 Students** | 23.40 | 7.00 | 11.40 | 63.20 | -12.00 | 56.20 |
| **150 Students** | 19.73 | 8.66 | 10.13 | 59.06 | -9.60 | 50.40 |
| **200 Students** | 26.30 | 10.30 | 11.30 | 58.90 | -15.00 | 48.60 |

*\* Simulation runtime = 9 minutes*

*1) Structure 1:* This configuration is a simplified baseline layout designed to model a typical school floor plan. It consists of long hallways intersected by multiple classrooms, with exits located at the far ends of the structure. The primary objective of this structure is to evaluate how the Gun-Detection System influences evacuation patterns in a standard, uncomplicated layout. By testing in this environment, we can better understand the system's ability to alert students to the shooter's presence, facilitate efficient evacuation routes, and minimize casualties.

*2) Gyte (Floor 1):* The Gyte floor structure represents the first floor of the PNW Gyte Building, with multiple interconnected hallways and classrooms [16]. Its design incorporates several exits, bottleneck points, and varied room layouts, providing a realistic scenario for examining how students evacuate or hide in response to an active shooter. The simulation on this floor structure allows for the evaluation of the Gun-Detection System's impact on evacuation success, as well as its influence on minimizing casualties in a more complex architectural environment.

*3) SULB (Floor 3):* The PNW's SULB floor 3 represents a more intricate layout, featuring a multi-level structure with increased spatial complexity [17]. This third-floor layout introduces additional challenges, such as stairwells, isolated rooms, and longer hallways, which affect both the shooter's movement and the student's ability to evacuate or hide. The inclusion of this structure provides insights into how varying floor designs can impact the effectiveness of the Gun-Detection System.

### D. Simulation Setup

To evaluate the effectiveness of the Gun-Detection System on the levels of casualties and the number of evacuated people, we performed experiments on 3 different floor structures mentioned in section III-C. For each of these floor layouts: *Structure 1, Gyte (Floor 1),* and *SULB (Floor 3)*, we performed 16 experiments, varying both the number of students present on the floor and the duration of the simulation.

To ensure a fair and accurate comparison between scenarios with and without the Gun-Detection System, we maintained consistent starting positions for both the students and the shooter across all simulations. This controlled setup allowed us to isolate the impact of the Gun-Detection System on the outcomes, providing reliable data on how it influences casualty rates and evacuation efficiency under various conditions. By varying the number of students and the time intervals of the simulations, we were able to evaluate the system's performance across a broad range of potential real-world situations, offering deeper insights into its operational effectiveness.

## IV. RESULTS AND DISCUSSION

Implementing the gun detection system within our agent-based model significantly improved evacuation efficiency and reduced casualty rates across various scenarios. We conducted multiple simulations to evaluate the system's effectiveness under different conditions, including varying occupancy levels, shooter behaviors, and environmental layouts. The results for the evacuation and casualty rates for three different floor structures (Structure 1, Gyte (Floor 1), SULB (Floor 3)) are mentioned in tables [II - V], [VI - IX], and [X - XIII] respectively.

- *Evacuation Efficiency:* The introduction of the gun detection system resulted in an average increase of evacuation efficiency by approximately 51.62%, 43.91%, and 49.31% for Structure 1, Gyte (Floor 1), and SULB (Floor 3) respectively when compared to scenarios without the system. Agents equipped with real-time threat information were able to make informed decisions about their escape routes, prioritizing paths that minimized exposure to the shooter. This dynamic adaptation improved the overall flow of evacuation and reduced instances of congestion, particularly in high-risk areas, leading to faster and safer evacuations.
- *Casualty Rates:* The integration of the gun detection system led to a significant decrease in casualty rates by approximately 10.09%, 6.99%, and 8.59% for Structure 1, Gyte (Floor 1), and SULB (Floor 3) respectively when compared to scenarios without the system. In the absence of the gun detection system, the average number of casualties was notably higher. However, with the gun detection system in place, agents were able to avoid the shooter's line of sight and adjust their routes based on real-time information. This resulted in a marked reduction in casualties, demonstrating that dynamic threat detection played a critical role in improving the overall safety of individuals during evacuation.

Comparative analysis for the evacuation and casualty rates for three different floor structures (Structure 1, Gyte (Floor 1), SULB (Floor 3)) are represented in figures 2, 3, and 4 respectively.

## V. CONCLUSION AND FUTURE WORK

In this research, we developed a multi-agent simulation that integrates a gun detection system to enhance evacuation efficiency during mass school shooter incidents. Our simulations demonstrated that incorporating gunshot detection significantly reduces evacuation time and reduces casualty rates compared to those without the system in place. By enabling agents to adapt their behaviors based on real-time information about the shooter's location, we established a more effective response framework for emergencies in school settings. The results highlight the limitations of existing evacuation strategies that do not account for dynamic threats and illustrate the importance of integrating technological solutions into emergency preparedness plans. Our findings provide a compelling case for the adoption of gun detection systems as part of comprehensive safety protocols in educational institutions. Building on the promising results of this study, future research can further refine and extend the model in several important ways. One key avenue is the incorporation of advanced machine learning techniques, such as reinforcement learning, to allow agents to adapt their behaviors dynamically in real time. This could enable continuous improvement in evacuation strategies as the model learns from evolving threats.

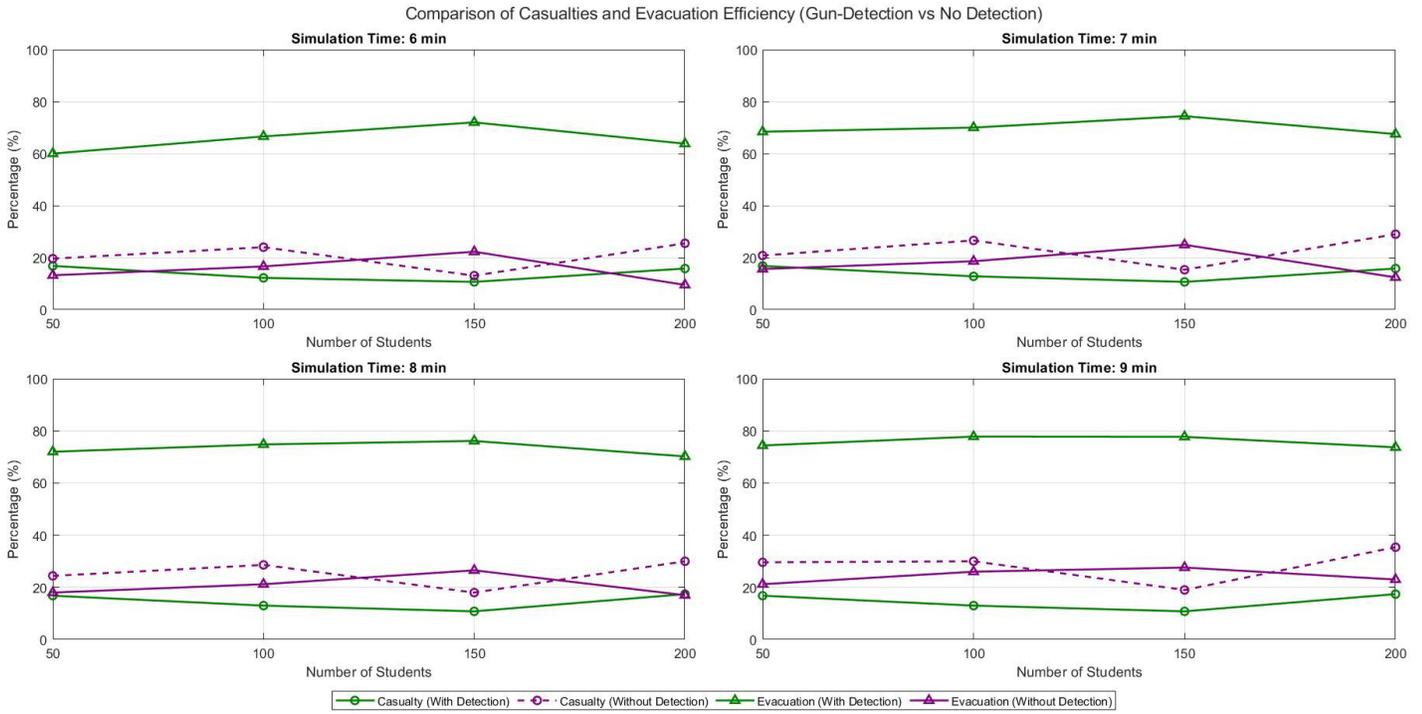

Fig. 2. Casualties and Evacuation Efficiency Comparison for Structure 1

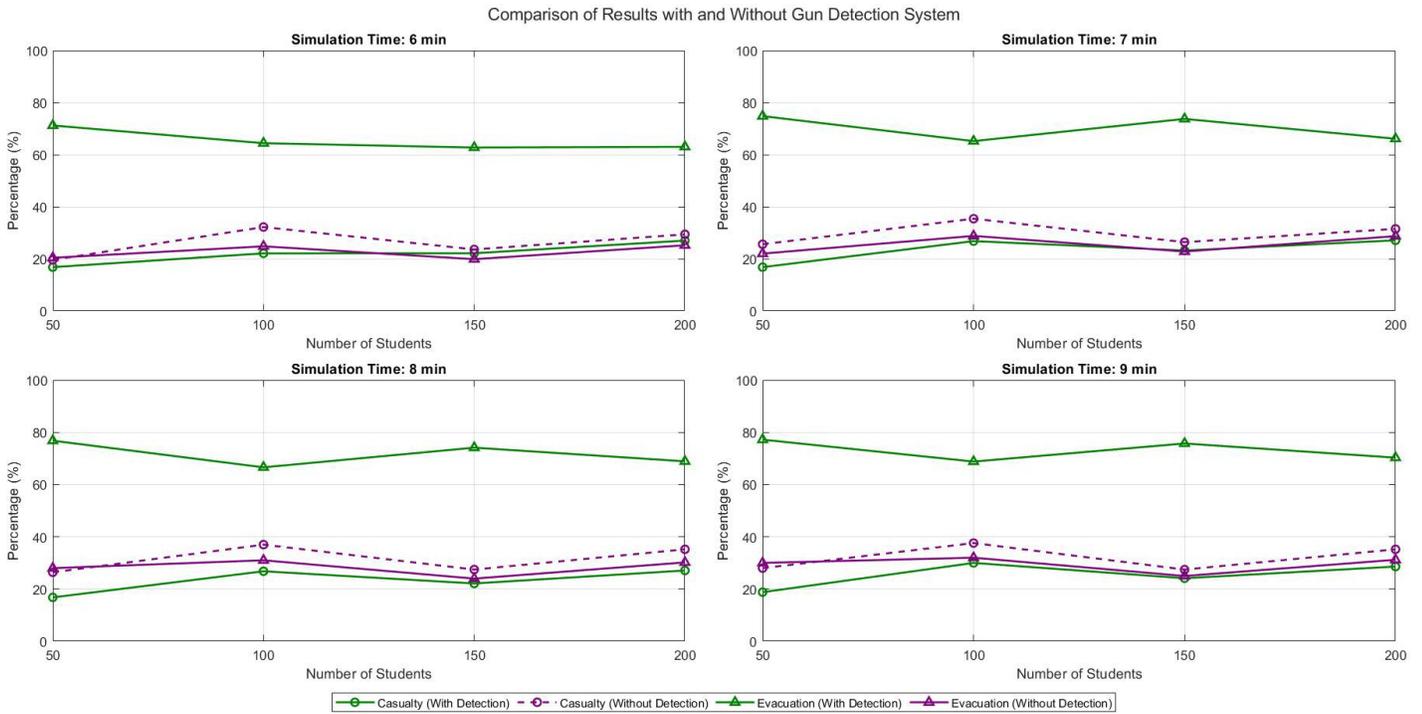

Fig. 3. Casualties and Evacuation Efficiency Comparison for Gyte (Floor 1)

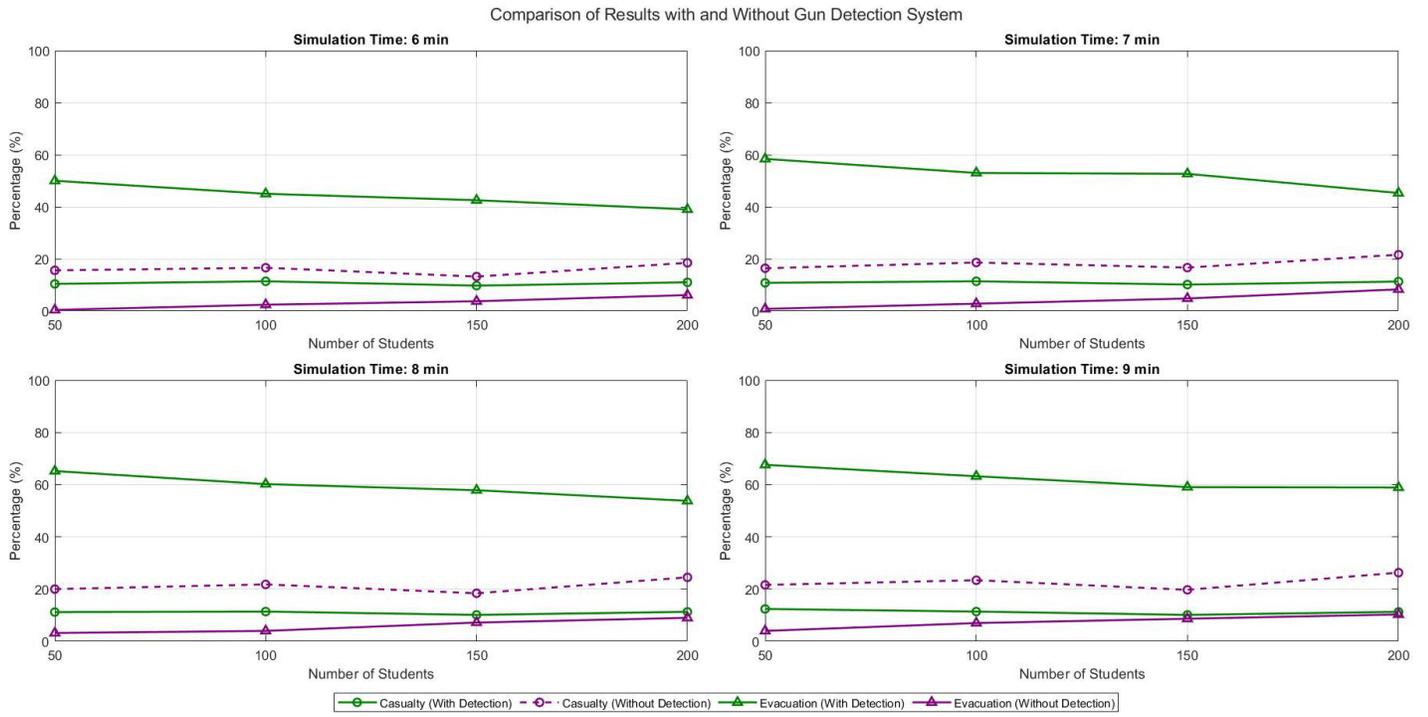

Fig. 4. Casualties and Evacuation Efficiency Comparison for SULB (Floor 3)

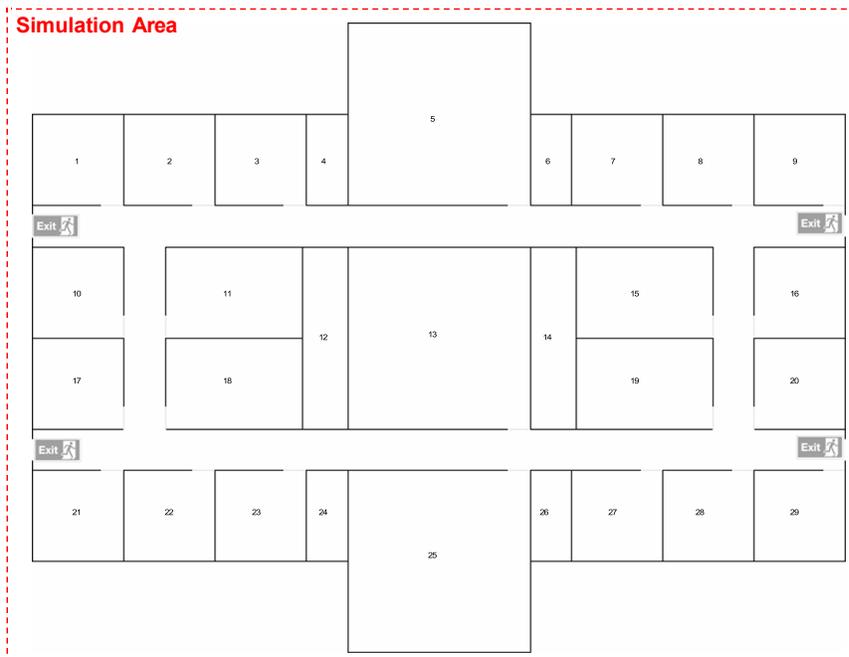

Fig. 5. Simulated map layout for Structure 1

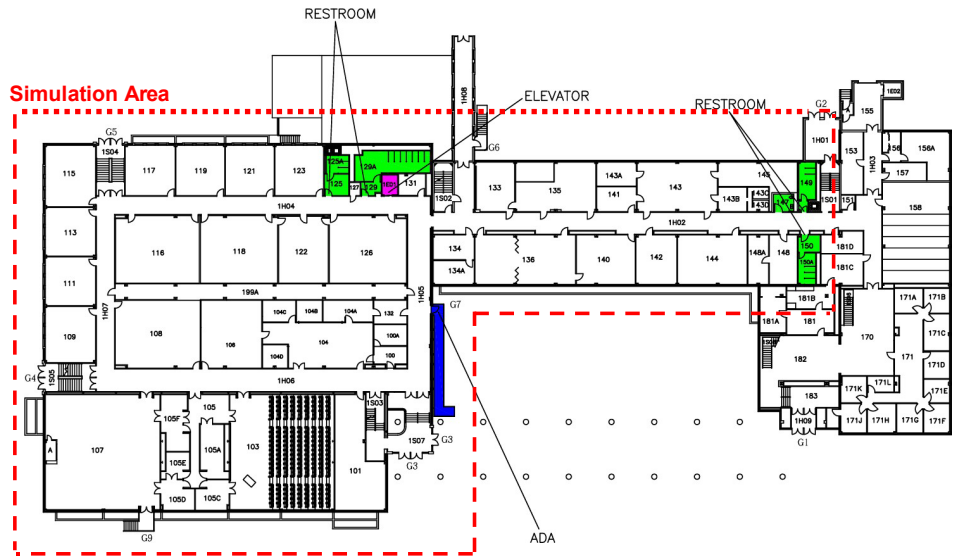

Fig. 6. Simulated map layout for PNW Gyte (Floor 1)

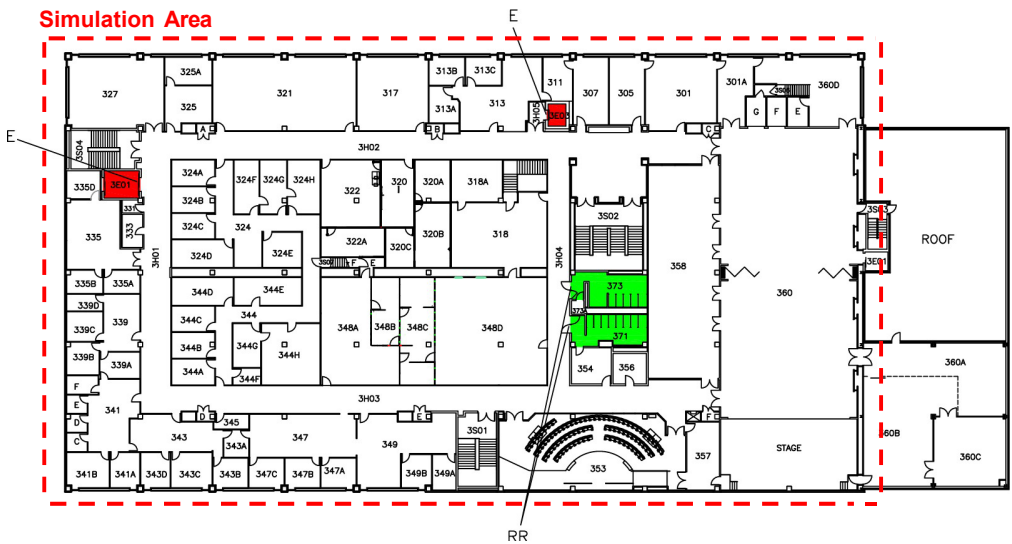

Fig. 7. Simulated map layout for PNW SULB (Floor 3)